\documentclass[a4paper,11pt]{article}
\usepackage{amsmath,amsthm,amssymb}
\usepackage{fullpage}
\usepackage{graphicx}

\newcommand{\x}{u}
\newcommand{\y}{v}
\newcommand{\be}{\begin{equation}}
\newcommand{\ee}{\end{equation}}
\begin{document}

\title{Randomness vs Non Locality and Entanglement}

\author{Antonio Ac\'{\i}n$^{1}$, Serge Massar$^2$, Stefano Pironio$^2$\\[0.5em]
\it\small{$^1$ICFO-Institut de Ciencies Fotoniques, 08860 Castelldefels (Barcelona), Spain.}\\
\it\small{and ICREA-Institucio Catalana de Recerca i Estudis Avan\c cats, 08010 Barcelona, Spain.}\\
\it\small{$^3$ Laboratoire d'Information Quantique, Universit{\'e}  Libre de Bruxelles, Belgium.}
}

\maketitle

\begin{abstract}
According to quantum theory, the outcomes obtained by measuring an
entangled state necessarily exhibit some randomness if they
violate a Bell inequality. In particular, a maximal violation of
the CHSH inequality guarantees that 1.23 bits of randomness are
generated by the measurements.  However, by performing
measurements with binary outcomes on two subsystems one could in
principle generate up to two bits of randomness. We show that
correlations that violate arbitrarily little the CHSH inequality
or states with arbitrarily little entanglement can be used to
certify that close to the maximum of two bits of randomness are
produced. Our results show that non-locality, entanglement, and
the amount of randomness that can be certified in a Bell-type
experiment are inequivalent quantities. From a practical point of
view, they imply that device-independent quantum key distribution
with optimal key generation rate is possible using almost-local
correlations and that device-independent randomness generation
with optimal rate is possible with almost-local correlations and
with almost-unentangled states.
\end{abstract}

\maketitle

Two of the most remarkable features of quantum theory are its
intrinsic randomness and its non-local character. The conclusion
that measurements on quantum systems yield random results was
first reached by Born and is now one of the basic axioms of the
theory. The intuition that measurements on entangled quantum
systems give rise to correlations that exhibit some form of
non-locality was made precise by Bell, whose work led the way to
the introduction of a series of inequalities that must be
satisfied by any locally causal theories, but which are violated
by quantum theory \cite{ref:Bell}.

These two -- a priori independent -- properties of quantum theory
are related through a third one, the no-signalling principle. The
no-signalling principle states that the outcomes of measurements
on separated systems cannot be used to send instantaneous signals.
Any theory that satisfies the no-signalling principle and which is
non-local, is also necessarily intrinsically random
\cite{ref:valentini,bhk,agm}. More precisely, if the measurement
outcomes of a Bell-type experiment violate a Bell inequality, then
these outcomes cannot be perfectly predicted within a
no-signalling (hence within quantum) theory. This conclusion holds
independently of any hypothesis on the type of measurements
performed or on the quantum systems (it even hold in non-quantum
theories provided they satisfy no-signalling). Conversely, if no
Bell inequalities are violated in a Bell experiment, then the
experimental results admits a purely deterministic explanation if
no additional hypothesis are made on the underlying system
\cite{ref:fine}.

The quantitative aspects of this fundamental connection between
non-locality and randomness have hardly been explored. Here we
address this problem within the quantum formalism (i.e. we do not
look at post-quantum theories) and investigate the relation
between non-locality, entanglement (which is necessary to produce
non-local correlations), and the amount of randomness necessarily
present in a Bell experiment.

Beyond its fundamental interest, this question is also motivated
by the recent development of device-independent randomness
generation (DIRNG)~\cite{colbeck,nat} and quantum key distribution
(DIQKD)~\cite{bhk,MayersYao,acinprl}. The observation that the
outcomes of measurements performed on two separate quantum systems
are necessarily random if they violate a Bell inequality can be
exploited to certify the randomness of strings or the secrecy of
shared keys generated in quantum cryptographic protocols without
the necessity to model the quantum state or the measurement
devices, thereby notably increasing the security of such
protocols. The quantitative study of the relation between
entanglement, non-locality, and randomness allows to determine the
minimal resources needed for DIRNG and DIQKD.

We focus here mostly on the simplest case in which two
measurements with binary outcomes can be applied to each one of
two separated quantum systems. In this context, the only
facet-defining Bell inequality is the CHSH inequality
\cite{ref:chsh,pra}, whose violation is thus a necessary condition
to certify the presence of randomness. The amount of violation of
the CHSH inequality can also be considered a natural measure of
non-locality: it uniquely determines for instance the maximum
local weight of a given set of correlations or the minimal amount
of communication required to simulate them \cite{pironio}.
Naively, one would thus expect a direct relation between the
amount of CHSH violation, and therefore of entanglement, and the
randomness produced in a Bell-type experiment, i.e., the less CHSH
violation or entanglement, the less randomness.

Our analysis, however, show that this intuition is not correct and
that the relation between these three concepts is much subtler than expected. In our scenario, where two subsystems are
measured and where each measurement results in one out of two
possible outcomes, the maximal amount of ``local randomness"
characterizing an individual outcome is 1 bit, while the maximal
amount of ``global randomness" characterizing the joint pair of
outcomes is 2 bits. We introduce here non-local correlations that
are arbitrarily close to the local region (i.e, which violate
arbitrarily little the CHSH inequality) or that arise from states
with arbitrarily little entanglement, yet which necessarily imply
that (arbitrarily close to) the maximal amounts of local or global
randomness are generated each time the system is measured. To
obtain these results we partially characterize the boundary of the
set of quantum correlations by introducing a family of Bell
inequalities and by determining the quantum points that maximally
violate them, i.e., by computing their ``Tsirelson bounds''
\cite{ref:T}.

Before presenting our main results, we introduce the notation and
definitions that will be used in the remainder of the paper and
establish three useful technical facts.

\section{Notation and definitions}
\paragraph{Bell experiments.}
We consider measurements on two distinct systems, denoted A and B.
On system A, one of two possible measurements $\x\in\{1,2\}$ are
carried out, resulting in one of two possible outcomes
$a\in\{-1,1\}$. Similarly, measurements $\y\in\{1,2\}$ are carried
out on system B, yielding outcomes $b\in\{-1,1\}$. We denote
$P(ab|\x\y)$ the probability to obtain the pair of outcomes
($a,b$) when the measurement settings $(\x,\y)$ are used. We focus
here on quantum probabilities, i.e. we assume that the
distribution $P$ is of the form
\begin{equation}
P(ab|\x\y)=\text{tr}[M_{a|\x}\otimes M_{b|\y}\, \rho]\,,
\end{equation}
where $\rho$ is a quantum state in some arbitrary Hilbert space
$H_A\otimes H_B$ and $M_{a|\x}$ and $M_{b|\y}$ are measurement
operators, i.e., they are positive and sum to the identity on
$H_A$ and $H_B$, respectively. By increasing the dimension of the
Hilbert spaces $H_A$ and $H_B$, we can without loss of generality
assume the positive operators $M_{a|\x}$ and $M_{b|\y}$ to be
projections. The measurements on system A and B can thus be
described by hermitian observables $A_\x=M_{0|\x}-M_{1|\x}$ and
$B_\y=M_{0|\y}-M_{1|\y}$ with eigenvalues $\pm 1$. We say that the
state $\rho$ and the observables $M=\{A_{\x},B_\y\}$ form a
quantum realization $\{\rho,M\}$ for $P$. In term of the
expectation values of the measurements $A_\x$ and $B_\y$, the
probabilities $P(ab|\x\y)$ can be expressed as \be
P(ab|\x\y)=\frac{1}{4}\left(1+a \langle A_\x\rangle +b \langle
B_\y\rangle + ab \langle A_\x B_\y\rangle\right). \ee

A Bell inequality is a linear constraint
$I=\sum_{ab\x\y}I_{ab\x\y}P(ab|\x\y)\leq I_L$ on $P$ that is
satisfied by every locally causal distribution, but which can be
violated by quantum distributions. The bound $I_L$ is called the
local bound of the inequality. We say that $P$ is non-local if it
violates a Bell inequality, i.e., $I>I_L$. In our scenario (two
binary measurement per party), there is a unique (up to
relabelling of the measurement outcomes and settings) facet
inequality, the CHSH inequality \be I=\langle
A_{1}B_{1}\rangle+\langle A_{1}B_{2}\rangle+\langle
A_{2}B_{1}\rangle-\langle A_{2}B_{2}\rangle\leq 2. \ee That is, an
arbitrary Bell inequality is violated only if the CHSH inequality
is also violated. Furthermore, the amount of violation $I>I_L$ of
the CHSH inequality can be viewed as a proper measure of the
non-locality of a given distribution $P$: for instance the optimal
amount of average communication $C$ required to simulate
classically a non-local distribution is directly related to the
CHSH violation through $C=I/2-1$ \cite{pironio}.

\paragraph{Randomness.}
Within the quantum formalism, two types of randomness have to be
distinguished: the genuine, intrinsic randomness of pure states
and the randomness of mixed states, which merely represents a lack
of knowledge about the definite state of the system. It is the
first type of randomness that we want to characterize here.

We quantify the randomness of the outcome pair $(a,b)$ resulting
from the measurement of the observables $A$ and $B$ on a given
pure state $|\psi\rangle\in H_A\otimes H_B$ through the
\emph{guessing probability}
\begin{equation}\label{gpure}
G(\psi,A,B)=\max_{ab    } P(ab|\psi,A, B)\,.
\end{equation}
where $P(ab|\psi,A,B)=\frac{1}{4}\left(1+a \langle A\rangle_\psi
+b \langle B\rangle_\psi + ab \langle A B\rangle_\psi\right)$ are
the corresponding joint outcome probabilities. The quantity
(\ref{gpure}) corresponds to the probability to guess correctly
the outcome pair $(a,b)$, since the best guess that one can make
is simply to ouput the most probable pair. The guessing
probability can be expressed in bits and is then known as the
\emph{min-entropy}  $H_\infty(\psi,A,B)=-\log_{2}G(\psi,A,B)$. If
a specific pair of outcomes $(a,b)$ is certain to occur, then the
guessing probability takes its maximal value 1 corresponding to 0
bits of min-entropy, while if all four possible pairs of outcomes
are equally probable, it takes its minimal value $1/4$
corresponding to $2$ bits of min-entropy.\\
\emph{Example:} Let
$\psi=(|00\rangle+|11\rangle)/\sqrt{2}$, $A=\sigma_z$,
$B=\sigma_x$. Then $P(ab|\psi,A,B)=1/4$ for all $a,b=0,1$, hence
$G(\psi,A,B)=1/4$.

For a mixed state $\rho\in B(H_A\otimes H_B)$, we define the
guessing probability associated to measurements $A$ and $B$ as
\begin{equation}\label{gmix}
G(\rho,A,B)=\max_{q_\lambda,\psi_\lambda} \sum_\lambda q_\lambda G(\psi_\lambda,A, B)\,.
\end{equation}
where the maximum is taken over all pure state decompositions
$\rho=\sum_\lambda q_\lambda
|\psi_\lambda\rangle\langle\psi_\lambda|$. This corresponds to the
maximal average guessing probability given the knowledge of which
underlying state $|\psi_\lambda\rangle$ in the ensemble has been
prepared. Equivalently, it corresponds to the maximal guessing
probability of someone that possess a quantum system correlated
with $\rho$ and who can perform measurements on his system to
guess the outcomes $(a,b)$.\\
\emph{Example:} Let
$\rho=(|00\rangle\langle 00|+|11\rangle\langle 11|)/2$,
$A=\sigma_z$, $B=\sigma_x$. Then $P(ab|\rho,A,B)=1/4$ for all
$a,b=0,1$ as above, but $G(\psi,A,B)=1/2$.

Our aim here is to analyze the fundamental constraints on the
randomness of a joint probability distribution $P(ab|uv)$ that
follow from its non-local properties alone, independently of any
particular quantum realization. Given a quantum distribution $P$,
we thus define the (realization-independent) guessing probability
of the outcome pair $(a,b)$ associated to the measurement choices
$(\x,\y)$ as
\begin{equation}\label{g}
G\left(P,\x,\y\right)=\max_{\{\rho,M\}\to P} G(\rho,A_\x,B_\y)
\end{equation}
where the maximum is taken over all quantum realisations
$\{\rho,M\}$ compatible with $P$, i.e., satisfying
$P(ab|\x\y)=\text{tr}[M_{a|\x}\otimes M_{b|\y}\, \rho]$.\\
\emph{Example:} Let $P$ be the quantum distribution that arises
from the measurements $A_1,B_1=\sigma_z$, $A_2=B_2=\sigma_x$ on
the state $\psi=(|00\rangle+|11\rangle)/\sqrt{2}$. Then we have
$P(ab|1,2)=1/4$ for all $a,b=0,1$, as in the two examples above.
However $G(P,1,2)=1$ since we can reproduce the entire
distribution $P$ by measuring the
$\mathbb{C}^4\otimes\mathbb{C}^4$ state
$\rho_{AB}=\frac{1}{4}\sum_{z_0,z_1=0}^1 (|z_0z_1\rangle\langle
z_0z_1|)_A\otimes(|z_0z_1\rangle\langle z_0z_1|)_B$ with the
observables $A_1=B_1=\sigma_z\otimes I$, $A_2=B_2=I\otimes
\sigma_z$.

In the same way as above, we can also define the
(realization-independent) guessing probability
$G\left(P,\x\right)$ of the single outcome $a$ associated to the
measurement choice $\x$, which has corresponding min-entropy
comprised between $0$ and $1$ bits.  In the following, we will
often write $G_{uv}$ and $G_u$ for $G(P,u,v)$ and $G(P,u)$ to
shorten the notation.

\paragraph{Randomness and non-locality.}
A distribution $P$ is said to be local deterministic if a
measurement of $\x$ always return an outcome $a=\alpha_\x$ and a
measurement of $\y$ always return an outcome $b=\beta_\y$, i.e.,
if $P(ab|\x\y)=\delta(a,\alpha_\x)\delta(b,\beta_\y)$.  Clearly a
local deterministic distribution admits a pure-state quantum
realization with guessing probability 1 (take for instance
$|\psi\rangle =
|\alpha_1,\alpha_2\rangle\otimes|\beta_1,\beta_2\rangle$,
$A_\x=\sum_{\alpha_\x} \alpha_\x|\alpha_\x\rangle\langle
\alpha_\x|$ and similarly for $B_\y$). Since a distribution is
local if and only if it can be written as a convex sum of local
deterministic distributions~\cite{ref:fine}, the violation of a
Bell inequality is a necessary condition for the guessing
probabilities $G_{uv}$ and $G_u$ to be different from 1.
Furthemore, it is also a sufficient condition, since non-local
correlations cannot be reproduced deterministically in quantum
theory~\cite{ref:valentini,bhk,agm}. The guessing probabilities
$G_{uv}$ and $G_u$ are thus different from 1 if and only if $P$
violates a Bell inequality, that is in our scenario, if and only
if it violates the CHSH inequality.

In general, for any given Bell inequality, one can derive bounds
$G_{uv}\leq f_{uv}(I)$ and $G_u\leq g_u(I)$ on the guessing
probabilities as a function of the amount of Bell
violation~$I$~\cite{nat}. Here, we will characterize the amount of
randomness associated to the Bell expressions
\begin{equation}
{I}^\beta_{\alpha}=\beta \langle A_1\rangle+ \alpha\langle
A_{1}B_{1}\rangle+\alpha\langle A_{1}B_{2}\rangle+\langle
A_{2}B_{1}\rangle-\langle
A_{2}B_{2}\rangle\,,\label{eq:Iab}\end{equation} which depend on
two parameters $\alpha$ and $\beta$. Without loss of generality,
we assume that $\alpha\geq 1$ and  $\beta\geq 0$ (the expressions
where either $\alpha<1$ or $\beta<0$ can be shown to be equivalent
to the expressions with $\alpha\geq 1$ and $\beta\geq 0$ by
relabelling the measurement settings and outcomes). To simplify
the notation we denote by $I_\alpha$ the Bell expression
$I_\alpha=I_\alpha^0=\alpha\langle A_{1}B_{1}\rangle+\alpha\langle
A_{1}B_{2}\rangle+\langle A_{2}B_{1}\rangle-\langle
A_{2}B_{2}\rangle$. When $\alpha=1$, $I_\alpha$ coincides with the
CHSH expression. The local bound of $I_\alpha^\beta$ is easily
found to be $\beta+2\alpha$.

In the following, we will be interested in the maximal amount of
randomness that can in principle be guaranteed by the Bell
expressions $I_\alpha^\beta$, that is, we will be interested in
the guessing probabilities $G_{uv}$ and $G_u$ under the constraint
that $I_\alpha^\beta$ is maximally violated.

\section{Technical preliminaries}
We start by presenting three useful technical results.

\paragraph{Reduction to two dimensions.}
First, note that in our scenario (two observables with binary
outcomes per system), it is sufficient to restrict the analysis to
pure two-qubit states.  More precisely, let $G(\Psi,A_u,B_v)\leq
f_{uv}(I)$ and $G(\Psi,A_u)\leq g_u(I)$ be bounds on the guessing
probabilities that are satisfied by any pure two-qubit state
\begin{equation}
|\Psi\rangle=\cos\theta\ |00\rangle+\sin\theta\ |11\rangle\,\label{eq:system}
\end{equation}
and non-degenerate Pauli observables
\begin{equation}\label{obs}
{A_{\x}}=\vec{a}_\x\cdot\vec{\sigma}\,,\qquad {B_{\y}}=\vec{b}_\y\cdot\vec{\sigma}\,
\end{equation}
yielding a Bell violation $I$. In the above expressions, $\theta$
is an angle satisfying $0\leq \theta \leq \pi/4$,
$\vec{\sigma}=(\sigma_1,\sigma_2,\sigma_3)$ are the three Pauli
matrices, and $\vec{a}_\x=(a_\x^1, a_\x^2,a_\x^3)$ and
$\vec{b}_\y=(b_\y^1, b_\y^2,b_\y^3)$ are unit vectors. Without
loss of generality suppose that the functions $f_{uv}(I)$ and
$g_u(I)$ are concave (if not take their concave hull). Then for
any quantum distribution $P$ with Bell violation $I$, it holds
that $G_{uv}\leq f_{uv}(I)$ and $G_u\leq g_u(I)$.

To show this, we recall the well-known fact that in our scenario
any distribution $P$ arising by measuring a state $\rho\in
B(H_A\otimes H_B)$, where the dimensions $\dim(H_A)$ and
$\dim(H_B)$ are in principle arbitrary, can always be expressed as
a convex combination $P=\sum_{c} p_{c} P_c$ of distributions $P_c$
arising from measurements on systems with $\dim(H_A)\leq 2$ and
$\dim(H_B)\leq 2$ \cite{acinprl,ref:Masanes}. Further, by
convexity, it is sufficient to consider the case where each $P_c$
arise from measuring a pure state. Note that if either
$\dim(H_A)=1$, or $\dim(H_B)=1$, or one of the operators $A_\x$ or
$B_\y$ is degenerate (e.g. $A_\x=\pm I$), then the corresponding
point $P_c$ is necessary local; but any local distribution can be
expressed as a convex combination of points obtained by measuring
the state $|00\rangle$ with $\pm \sigma_z$ observables. It is
therefore completely general to assume that each $P_c$ admits a
realization in term of a pure two-qubit state, which can always be
written as (\ref{eq:system}) in the Schmidt basis, and
measurements corresponding to  non-degenerate Pauli observables of
the form (\ref{obs}).

Now let $\psi$ be an arbitrary (not necessarily two-qubit) pure
state and $A_u$, $B_v$ observables yielding a violation I. We then
have using the above observation that
$G(\psi,A_u,B_v)=\max_{ab}P(ab|\psi,A_u,B_v)=\max_{ab}\sum_c p_c
P_c(ab|uv)\leq \sum_c p_c \max_{ab} P_c(ab|uv)=\sum_c p_c
\linebreak[4] G(\psi_c, A_{u,c},B_{v,c})\leq \sum_c p_c f(I_c)\leq
f_{uv}(\sum_c p_c I_c)\leq f_{uv}(I)$, where we have expressed the
probabilities $P(ab|\psi,A_u,B_v)$ as a convex sum of
probabilities $P_c(ab|uv)$ arising from pure two-qubit states
$\psi_c$ and non-degenerate Pauli observables $A_{u,c}$, $B_{v,c}$
in the second equality, have used the bound on the guessing
probability valid for pure two-qubit states in the second
inequality, and the concavity of the function $f_{uv}$ in the
third inequality. Since the bounds $G(\psi,A_u,B_v)\leq f_{uv}(I)$
hold for any pure state, it follows from the definitions
(\ref{gmix}) and (\ref{g}) and again the concavity of $f_{uv}$
that $G_{uv}\leq f(I)$ hold for any distribution $P$. The same
reasoning applies to $G_{u}$.

\paragraph{Bound on predictability.}
Second, note that by measuring the state (\ref{eq:system}) with
the observables (\ref{obs}), one necessarily has
\begin{equation}\label{bias}
-\cos2\theta\leq \langle A_\x\rangle \leq \cos2\theta\,,
\end{equation}
the extremal values being obtained when $A_{\x}=\pm\sigma_{z}$.
One finds similarly $-\cos2\theta\leq \langle B_\y\rangle \leq
\cos2\theta$.

\paragraph{Optimal violation of $I_\alpha$ for $2\times 2$ systems.}
Finally, for any set of measurements (\ref{obs}) performed on the
state (\ref{eq:system}), the following inequality necessarily
holds
\begin{equation}
I_{\alpha}\leq 2\sqrt{\alpha^{2}+\sin^{2}2\theta}\,.\label{eq:Itheta}
\end{equation}
Furthermore, if $\theta>0$ there are only two probability
distributions $P$ saturating this inequality defined by the
expectation values
\begin{eqnarray}
&&\langle A_1\rangle=\pm \cos2\theta\,,\qquad     \langle A_2\rangle=0\,,\nonumber\\
&& \langle B_1\rangle=\langle B_2\rangle= \pm\cos\mu\cos2\theta\,,\nonumber \\
&&\langle A_1B_1\rangle =\langle A_1B_2\rangle= \cos\mu\,,\label{corr}\\
&&\langle A_2B_1\rangle=-\langle A_2B_2\rangle=\sin2\theta\sin\mu\,,\nonumber
\end{eqnarray}
where $\tan\mu=\sin2\theta/\alpha$. These two points are obtained
using the Pauli observables
\begin{eqnarray}
A_1&=&\pm\sigma_z\,,\nonumber\\
A_2&=&\cos \varphi\, \sigma_x + \sin\varphi\, \sigma_y\,,\nonumber\\
B_1&=&\pm \cos\mu\,\sigma_z+\sin\mu\,(\cos \varphi\, \sigma_x - \sin\varphi\, \sigma_y)\,,\nonumber\\
B_2&=&\pm \cos\mu\,\sigma_z-\sin\mu\,(\cos \varphi\, \sigma_x - \sin\varphi\, \sigma_y)\,,\label{optimalmst}
\end{eqnarray}
where $\varphi\in[0,2\pi[$ is a free parameter.

To show this, it is convenient to rewrite the state
(\ref{eq:system}) as $\rho=|\Psi\rangle\langle\Psi|$ with
\begin{equation*}
\rho=\frac{I}{4}+\cos2\theta\,\frac{\sigma_{z}\otimes
I}{4}+\cos2\theta\,\frac{I\otimes\sigma_{z}}{4}+\sum_{ij}T_{ij}\frac{\sigma_{i}\otimes\sigma_{j}}{4}\,,
\end{equation*}
where the $3\times 3$ real matrix $T$ has components
 \begin{equation*}
T_{xx}=\sin2\theta\ ,\ T_{yy}=-\sin2\theta\ ,\ T_{zz}=1,\ T_{ij}=0 \text{ for } i\neq j\,.
\end{equation*}
Following the method exposed in \cite{ref:H3}, we introduce two
normalised mutually orthogonal vectors $\vec{c}_1$ and $\vec{c}_2$
by \begin{equation} \vec{b}_1+\vec{b}_2=2\cos\mu\ \vec{c}_1\ ;\
\vec{b}_1-\vec{b}_2=2\sin\mu\
\vec{c}_2,\label{eq:cc'}\end{equation}
 where $\mu\in[0,\frac{\pi}{2}]$. We can then write
 \begin{equation}
I_{\alpha}=2\alpha\cos\mu\ (\vec{a}_1\cdot T\,\vec{c}_1)+2\sin\mu\
(\vec{a}_2\cdot T\,\vec{c}_2)\,.
\end{equation}
Let us now maximise $I_{\alpha}$ over all measurements
$A_{1},A_{2},B_{1},B_{2}$, while keeping the state (i.e. $T$)
fixed. We find \begin{eqnarray}
\max_{\vec{a}_1\vec{a}_2\vec{c}_1\vec{c}_2\mu} I_{\alpha} & = & \max_{\vec{c}_1\vec{c}_2\mu}\ 2\alpha\cos\mu\ |T\,\vec{c}_1|+2\sin\mu\ |T\,\vec{c}_2|\nonumber \\
 & = & \max_{\vec{c}_1\vec{c}_2}\ 2\sqrt{\alpha^{2}|T\vec{c}_1|^{2}+|T\vec{c}_2|^{2}}'\,,\label{eq:maxIalpha}\end{eqnarray}
where the first equality obtains when
$\vec{a}_u=T\,\vec{c}_u/|T\,\vec{c}_u|$
and the second equality when $\tan\mu=\sin2\theta/\alpha$.
 Since $\vec{c}_1$ and $\vec{c}_2$ are orthogonal, and since $\alpha>1$,
the maximum of (\ref{eq:maxIalpha}) is  obtained when
$\vec{c}_1=\pm\vec{1}_{z}$ lies along the direction of the largest
eigenvalue of $T$, and $\vec{c}_2$ lies in the $x,y$ plane along
some arbitrary direction $\varphi$. We thus find that for the
state (\ref{eq:system}), $I_{\alpha}\leq
2\sqrt{\alpha^{2}+\sin^{2}2\theta}$, where this inequality is
saturated using measurements given in eq. (\ref{optimalmst}). Such
measurements yield the expectation values~(\ref{corr}) for any
value of the free parameter $\varphi$.

\section{Results}
\paragraph{Arbitrarily high randomness from arbitrarily low non-locality.}
From (\ref{eq:Itheta}), we deduce that the maximal quantum
violation of the $I_\alpha$ inequality is  $2\sqrt{\alpha^2+1}$
and that it can be obtained by measuring a maximally entangled
state, i.e. $\theta=\pi/4$ in (\ref{eq:system}). Further note from
(\ref{corr}) that there exists a unique pure two-qubit quantum
probability distribution achieving this maximum defined by
\begin{equation}\label{ia}\begin{split}
&\langle A_\x\rangle=\langle B_\y\rangle=0\\
&\langle A_1B_\y\rangle = \frac{\alpha}{\sqrt{1+\alpha^2}},\quad \langle A_2B_\y\rangle = \frac{(-1)^\y}{\sqrt{1+\alpha^2}}
\end{split}\end{equation}
By the convex reduction of general quantum distribution to
two-qubit distributions, this probability distribution is actually
the unique quantum distribution reaching the maximal quantum value
$2\sqrt{\alpha^2+1}$. The guessing probabilities $G_{uv}$ and
$G_{u}$  at the point of maximal violation thus simply correspond
to the guessing probabilities of the distribution (\ref{ia}).

In the case $\alpha=1$, we recover the well known properties of
the CHSH expression. It is bounded by $I_1\leq2\sqrt{2}$ (the
Tsirelson bound) and at the maximum the measurements $A_\x$ and
$B_\y$ are locally completely uncertain, i.e. $G_{u}=G_v=1/2$.
While Alice's  and Bob's outcomes are locally completely random,
they are not completely uncorrelated  and one finds
$G_{uv}=1/4+\sqrt{2}/8\simeq 0.427$, corresponding to
$-\log_2{G_{uv}}\simeq 1.23$ bits of global randomness in the pair
$(a,b)$.

Let us now consider the $I_{\alpha}$ inequality with $\alpha>1$.
As in the CHSH case, we see from (\ref{ia}) that at the point of
maximal violation, the outcomes of $A_\x$ and $B_\y$ are locally
completely uncertain, i.e. $G_{u}=G_v=1/2$.
Note, however, that when the $I_\alpha$ inequality is maximally
violated, the CHSH inequality has the value
$I_1=2(\alpha+1)/\sqrt{\alpha^2+1}\simeq 2+2/\alpha$, i.e., for
large $\alpha$, the CHSH violation is arbitrarily small. Thus we
see that perfect local randomness can be obtained with points that
are arbitrarily close to the local region.

A stronger results holds if we consider the global randomness.
From (\ref{ia}), we see that at the point of maximal violation of
the $I_\alpha$ inequality, the guessing probability
$G(P,2,v)=1/4\times (1+ 1/ \sqrt{\alpha^2+1})$ which is smaller
(i.e. it corresponds to more randomness) than the CHSH guessing
probability for any $\alpha>1$. Furthermore, for large $\alpha$,
we have  $-\log_2{G(P,2,\y)}\simeq 2-\ln(2)/\alpha$, which is
arbitrarily close to the optimal value of $2$ bits of global
randomness even though the quantum point becomes arbitrarily close
to the local region.

Note that in the case of the local randomness, we can characterize
the guessing probability $G_u$ not only for the point of maximal
violation but for any degree of violation. That is we obtain the
complete curve $G_u\leq g(I_\alpha)$, which takes the form
\begin{equation}
G_u \leq \frac{1}{2} +\frac{1}{2}\sqrt{1 + \alpha^2
-\frac{I_\alpha^2}{4}} \,.\label{eq:boundPx}
\end{equation}
This bound is tight in the case $\x=1$. It was derived previously
in the CHSH case ($\alpha=1$) for qubits (i.e. not in a
device-independent way) in \cite{ref:JB}, and in the
device-independent scenario in \cite{nat,DIQKD}.

To show (\ref{eq:boundPx}), remember that it is sufficient to
establish this relation for the case of pure two-qubit
states~(\ref{eq:system}), as discussed previously. From
Eq.~(\ref{eq:Itheta}), it follows that the only states compatible
with a given value of $I_\alpha$, are those satisfying
$\cos2\theta\leq \sqrt{1+\alpha^ 2-{I_\alpha^2}/{4}}$. Using this
inequality in (\ref{bias}),  we obtain (\ref{eq:boundPx}). Note
that this bound is tight in the case $\x=1$, since the
correlations (\ref{corr}) saturating (\ref{eq:Itheta}) for fixed
$\theta$, also saturate (\ref{bias}).

\textbf{Perfect local randomness from any partially entangled
state.} Let us now characterize the point of maximal violation of
the inequality $I_\alpha^\beta$ with $\beta>0$. As usual, it is
sufficient to consider pure-state two-qubit correlations.
Combining Eq. (\ref{bias}) and (\ref{eq:Itheta}), we find the
inequality $\langle A_1\rangle \leq
\sqrt{1+\alpha^2-I_\alpha^2/4}$. Inserting this bound for $\langle
A_1\rangle$ in (\ref{eq:Iab}), we obtain $I_{\alpha}^\beta\leq
I_\alpha+\beta\sqrt{1+\alpha^2-I_\alpha^2/4}$. This expression is
easily seen to be maximized when
$I_\alpha=2\sqrt{1+\alpha^2}/\sqrt{1+\beta^2/4}$, implying that
the maximal quantum violation of the inequality $I_{\alpha}^\beta$
is $2\sqrt{(1+\alpha^2)(1+\beta^2/4)}$. Furthermore, provided that
$\beta\neq 2/\alpha$, the inequality $\langle A_1\rangle \leq
\sqrt{1+\alpha^2-I_\alpha^2/4}$ that we used in the derivation of
this bound is uniquely saturated by the quantum point (\ref{corr})
(with the $+$ sign) and thus we necessarily have at the point of
maximal violation that $\langle A_2\rangle =0$, i.e. $G_2=1/2$ for
any $\beta\neq 2/\alpha$. Note further that the maximal violation
is obtained by measuring the state (\ref{eq:system}) with $\theta$
such that
$I_\alpha=2\sqrt{1+\alpha^2}/\sqrt{1+\beta^2/4}=2\sqrt{\alpha^2+\sin^22\theta}$,
or $\sin2\theta=\sqrt{(1-\alpha^2\beta^2/4)/(1+\beta^2/4)}$.
Taking for instance $\alpha=1$ and $0<\beta<2$ therefore implies
that 1 bit of local randomness can be certified by the violation
of a Bell inequality for any partially entangled state.

\textbf{Arbitrarily high global randomness from almost unentangled
states.} We showed above that using the $I_\alpha$ inequality in
the limit $\alpha\to 0$ one can certify that arbitrarily close to
2 bits of randomness are produced from a $2\times 2$ maximally
entangled quantum system. We now show that one can also certify
that arbitrarily close to 2 bits of randomness are produced in the
limit where the $2\times 2$ system tends towards an unentangled
state.

To this end, we consider a slightly more complex situation than
the one analyzed so far, in which  Alice and Bob each have four
two-outcome measurements $A_1,A_2,A'_1,A'_2$ and
$B_1,B_2,B'_1,B'_2$ (since the measurements have binary outcomes,
the maximal randomness associated to a pair of joint measurements
is still 2 bits). Let ${I}_{\alpha}^\beta$ and
${I'}_{\alpha}^\beta$ denote the Bell expression (\ref{eq:Iab})
obtained using the unprimed measurements $A_\x, B_\y$, or the
primed measurements $A'_\x, B'_\y$ respectively, where for
${I'}_{\alpha}^\beta$ the roles of Alice and Bob are reversed.
Suppose that both $I_{\alpha}^\beta$ and ${I'}_{\alpha}^\beta$ are
maximally violated, i.e.,
$I_{\alpha}^\beta={I'}_{\alpha}^\beta=2\sqrt{(1+\alpha^2)(1+\beta^2/4)}$.
To determine the corresponding guessing probability $G_{uv}$, it
is clearly sufficient to characterize the maximal guessing
probability $G(\Psi,A_u,B_v)$ for all pure states $\Psi$ and all
observables for wich both $I_{\alpha}^\beta$ and
${I'}_{\alpha}^\beta$ are maximally violated. From the previous
results, we know (provided that $\beta\neq 2/\alpha)$ that the
outcomes of $A_2$ and $B'_2$ are locally completely random. i.e.,
$\langle A_2\rangle_\Psi=\langle B'_2\rangle_\Psi=0$. We now show
that the results of $A_2$ and $B'_2$ are almost not correlated for
$\beta\rightarrow 2/\alpha$, more precisely we show that $|\langle
A_2B'_2\rangle_\Psi|=\sqrt{(1-\alpha^2\beta^2/4)/(1+\beta^2/4)}$
for all $|\Psi\rangle$. If we take, e.g., $\alpha=1$,
$\beta=2-\epsilon$, this implies that
$G(\Psi,A_2,B'_2)=(1+|\langle A_2\rangle_\Psi|+|\langle
B'_2\rangle_\Psi|+|\langle A_2B'_2\rangle_\Psi|)/4\simeq
1/4+1/4\sqrt{\epsilon/2}$ and thus $G_{22'}\simeq
1/4+1/4\sqrt{\epsilon/2}$. Moreover, a maximal violation of both
$I_{\alpha}^\beta$ and ${I'}_{\alpha}^\beta$ can be obtained by
measuring a state (\ref{eq:system}) with $\sin2\theta\simeq
\sqrt{\epsilon/2}$. We thus see that arbitrarily close to $2$ bits
of global randomness can be certified using states that are almost
unentangled.

The intuition behind this result is that in the limit $\theta\to
0$ the state tends to $|\Psi\rangle\to|00\rangle$ while $A_2$ and
$B'_2$ are both measurements in the $x,y$ plane (see eq.
(\ref{optimalmst})). Hence the the corresponding measurement
outcomes become uncorrelated random bits in the limit. The main
difficulty is that since this Bell experiment involves four
measurements per party, we cannot directly reduce the analysis to
two-qubit states.

However we can simultaneously block-diagonalise in blocks of size
$2$ the pairs of observables $A_1, A_2$ and  $B_1, B_2$, as well
as the pairs of observables $A'_1, A'_2$,  $B'_1, B'_2$
\cite{ref:Masanes,acinprl}. Consider first the
block-diagonalisation of the pairs of observables $A_1, A_2$ and
$B_1, B_2$. Let $\{| 0_i\rangle,| 1_i\rangle\}$ $(i=1,\ldots,N_A)$
denote a basis for the $N_A$ blocks on Alice's side and similarly
let $\{| 0_j\rangle,|1_j\rangle_j\}$ $(j=1,\ldots,N_B)$ denote a
basis for the $N_B$ blocks on Bob's side. In each $2\times 2$
block, the unique state maximally violating $I_{\alpha}^\beta$ is
(up to local unitaries) the state
$|\psi_{\theta}\rangle_{ij}=\cos\theta |0_i0_j\rangle +\sin\theta
|1_i1_j\rangle$ where $\sin
2\theta=\sqrt{(1-\alpha^2\beta^2/4)/(1+\beta^2/4)}$. Since we
suppose that the global state $|\Psi\rangle$ violates maximally
this inequality, it necessarily has the form
$|\Psi\rangle=\sum_{ij}{c_{ij}} |\psi_{\theta}\rangle_{ij}$ where
$\sum_{ij}|c_{ij}|^2=1$ and the observables have the form
$A_1=\sum_i \sigma_z^i$ and $A_2=\sum_i (\cos \varphi_i \sigma_x^i
+ \sin \varphi_i \sigma_y^i)$ and $B_{1,2}=\sum_j
\cos\mu\sigma_z^j\pm\sin\mu(\cos \varphi_j \sigma_x^j -
\sin\varphi^j \sigma_y)$, i.e., they are sums of Pauli operators
acting on each block. The angles $\varphi_i$ and $\varphi_j$ are
not independent, since if $c_{ij}\neq 0$ is non zero they must be
equal. Therefore, for all $i,j$ for which $c_{ij}\neq 0$, one can
bring $\varphi_i=\varphi_j=0$ by simultaneous rotation of Alice's
basis and Bob's basis around the $z$ axis by opposite angles. From
now on we assume this is the case.

The matrix $c_{ij}$ has a singular value decomposition
$c_{ij}=\sum_k U_{ik} c_k V_{kj}$ where $U_{ik}$ and $V_{kj}$ are
unitary matrices and the $c_k\geq 0$ are non-negative real
numbers. We can therefore rewrite the state as
\begin{equation}
|\Psi\rangle=\sum_{k}{c_{k}} \left(\cos\theta |\bar{0}_k\bar{0}_k\rangle +\sin\theta |\bar{1}_k\bar{1}_k\rangle \right)\,,
\label{psibar}
\end{equation}
where $|\bar{0}_k\rangle_A=\sum_i U_{ik}|{0}_i\rangle_A$,
$|\bar{1}_k\rangle=\sum_i U_{ik}|{1}_i\rangle_A$, and
$|\bar{0}_k\rangle_B=\sum_j V_{jk}|{0}_i\rangle_B$,
$|\bar{1}_k\rangle_B=\sum_j V_{jk}|{1}_i\rangle_B$ (here and below
we add the index $A$ or $B$ whenever distinguishing between Alice
and Bob's states is not implicit from the notation). Importantly,
it is easily checked that the operators $A_1,A_2,B_1,B_2$ have the
same form expressed in the new basis as in the old basis. In
particular $A_1=\sum_k \bar{\sigma}_z^k$ and $A_2=\sum_k
\bar{\sigma}_x^k$.

Let us now apply the same operations to the block diagonalisation
of the pairs of observables $A'_1, A'_2$,  $B'_1, B'_2$. We can
bring the state to the form
\begin{equation}
|\Psi\rangle=\sum_{k}{c_{k}} \left(\cos\theta |\bar{0}'_k\bar{0}'_k\rangle +\sin\theta |\bar{1}'_k\bar{1}'_k\rangle \right)
\label{psibarprime}
\end{equation}
with the operators $B'_1=\sum_k \bar{\sigma}'^k_z$ and
$B'_2=\sum_k \bar{\sigma}_x'^k$. Note that the coefficients $c_k$
are the same and have the same degeneracy in eqs. (\ref{psibar})
and (\ref{psibarprime}) since the state is written in the Schmidt
basis. However the basis states $|\bar{0}_k\rangle_{A,B}$,
$|\bar{1}_k\rangle_{A,B}$ and $|\bar{0}'_k\rangle_{A,B}$,
$|\bar{1}'_k\rangle_{A,B}$  may differ. If the singular value
$c_k$ is non degenerate they may differ by a phase, whereas if the
singular value $c_k$ is degenerate they may differ by unitary
transformations.

Let us assume that the singular value $c_k$ has degeneracy $d$.
From now on we work within this $2d\times 2d$ block, and drop the
index $k$ (indeed all the operators commute with the projections
onto these $2d\times 2d$ blocks). We thus consider the normalised
state
\begin{eqnarray}
|\Psi\rangle&=&\frac{1}{\sqrt{d}}\sum_{l=1}^d \left( \cos\theta  |0_l\rangle|0_l\rangle + \sin \theta |1_l\rangle|1_l\rangle\right)\,\label{Psi1}\\
&=& \frac{1}{\sqrt{d}}\sum_{l=1}^d \left( \cos\theta  |0'_l\rangle|0'_l\rangle + \sin \theta |1'_l\rangle|1'_l\rangle\right)\,\label{Psi2}\end{eqnarray}
and measurements $A_2=\sum_l \sigma_x^l$ and $B'_2=\sum_l \sigma_x'^l$ (where for simplicity we have omitted the bar `` $\bar{\ }$ " over the states and operators). We can rewrite $|0'_l\rangle_{B} =\sum_m W_{lm}|0_m\rangle_{B}$ with $W$ the unitary matrix that transform from the $|0_l\rangle$ to the $|0'_l\rangle$ basis. We also write $|1'_l\rangle_{B} =\sum_m W_{lm}|1''_m\rangle_{B}$, where the states $|1''_m\rangle$ must be ortogonal to the $|0_l\rangle_B$ states, hence can be written as  $|1''_l\rangle_B=\sum_m R_{lm}|1_m\rangle_B$ for some unitary matrix $R$. It is easily verified using these relations that the operator $B'_2=\sum_l|0'_l\rangle\langle 1'_l|+|1'_l\rangle\langle 0'_l|=\sum_l|0_l\rangle\langle 1''_l|+|1''_l\rangle\langle 0_l|$. By computing explicitely the expectations value $\langle A_2B'_2\rangle_\Psi$ using this last expression for $B'_2$ on the state given by Eq.~(\ref{Psi1}), one finds $\langle A_2B'_2\rangle_\Psi=\frac{1}{d}\cos\theta\sin\theta\sum_l (R_{ll}+R_{ll}^*)=\sin(2 \theta)$, which is the desired result.
%
%

\section{Discussion}
In this work we have considered the relation between entanglement,
non locality, and the amount of randomness that can be certified
in a Bell-type experiment. These quantities are closely related:
entanglement is necessary for non locality, and non locality is
necessary for certifying randomness. The quantitative relations
between these concepts, however, are more subtle than expected. It
was already known that entanglement and non locality are
inequivalent resources~\cite{BGS,AGG,MS}. Here we have shown that
the amount of randomness that can be certified by a Bell-type
experiment is inequivalent to either of these two resources. 

Some understanding of why non-locality is inequivalent to
certified randomness can be obtained by going back to the
geometric picture. In our work, we have characterized part of the
boundary of quantum correlations and have shown that there exists
extremal quantum distributions $P(ab|uv)$ which are arbitrarily
close to the set of local correlations or which arise from
partially entangled states, yet which are close to uniformly
random for specific choices of $u$ and $v$, see Figure~1.
\begin{figure}\center
\includegraphics[width=120mm]{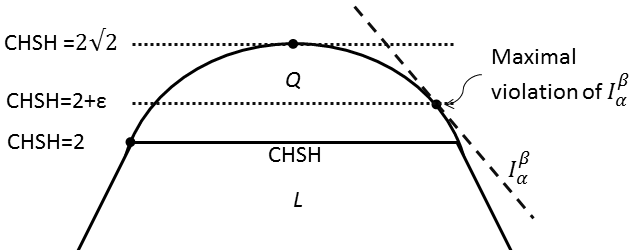}
\caption{Schematic representation of the set of local correlations
($L$) and quantum correlations ($Q$). A Bell expression defines a
hyperplane in the space of correlations. The CHSH hyperplane
$CHSH=2$ separates the local region from the non-local quantum
region. When the Bell inequalities $I_\alpha^\beta$ are maximally
violated, the corresponding hyperplanes become tangent to the
quantum boundary and identify one of the extremal points of $Q$.
The distribution $P$ associated to such extremal points may be
close to uniformly random for certain values of $u,v$, but need
not be highly non-local as measured by the CHSH violation (as
represented on the figure) or need not originate from maximally
entangled states.}
\end{figure}
This suggests that while non-locality is necessary to certify the
presence of randomness, its quantitative aspects are related to
the extremality of non-local correlations. In this sense, our work
goes in the same direction as \cite{FFW}, where extremality was
identified as a key property for the security of DIQKD.

From a practical point of view our results have direct
applications for DIRNG and DIQKD. The guessing probabilies
$G_{uv}$ and $G_u$ play a central role in the recent security
proofs for, respectively, DIRNG~\cite{nat} and
DIQKD~\cite{DIQKD,ren}. Upper-bounds on these quantities, as a
function of the amount of violation of a Bell inequality, directly
translate into bounds on the amount of randomness generated in
DIRNG protocols and on the key rate of DIQKD protocols. It follows
in particular from our results that the CHSH inequality is not
optimal for DIRNG but that higher generation rates, up to the
optimal value of 2 bits per use of the system, can be obtained
using other inequalities, and that randomness generation rates
superior to 1 bit per use of the system are possible from any
partially entangled states. In the context of DIQKD, the fact that
1 bit of local randomness can be extracted from maximally
entangled states irrespectively of the amount of violation of the
CHSH inequality implies that DIQKD with an optimal asymptotic rate
of 1 bit of secret key per use of the system is possible using
correlations that are almost local.

From a fundamental point of view, it is interesting to compare our
results to those that can be established for post-quantum theories
limited only by the no signalling principle. In this case, the
geometry of the space of non-local correlations corresponding to
experiments involving two possible binary measurements on each
subystem is very simple since the unique extremal points are the
local deterministic correlations and the Popescu-Rohlich
boxes~\cite{pra}. This implies that the amount of certifiable
randomness is proportional to the violation of the CHSH
inequality, and reaches at most 1 bit when the CHSH violation is
equal to 4. On the other hand, in the quantum case the amount of
certifiable randomness can be arbitrarily close to the maximal
possible value of 2 bits, i.e., more randomness can be extracted
from the non-local correlations of quantum theory than it would be
possible in the most non-local theory compatible with
no-signalling. It would be interesting to investigate if they are
other no-signalling theories allowing for maximal certifiable
randomness.

Finally, we have shown that arbitrarily close to 2 random bits can
be certified by maximally entangled states, as well as by states
with arbitrarily little entanglement. We conjecture that this
value can be reached for any value of the entanglement (the
parameter $\theta$ in eq. (\ref{eq:system})). It would also be
interesting to understand whether measurements beyond the
projective case provide any advantage or whether two bits is the
maximum amount of randomness that can be certified by $2\times 2$
states.

{\bf Acknowledgements.} We thank A. Boyer de la Giroday for
contributions during early stages of this project. This work was
supported by the ERC starting grant PERCENT, the European EU FP7
Q-Essence and QCS projects, the Spanish FIS2010-14830 and
Consolider-Ingenio QOIT projects, the Interuniversity Attraction
Poles Photonics@be Programme (Belgian Science Policy), the
Brussels-Capital Region through a BB2B Grant.

\end{document}